\documentclass[aps,prl,twocolumn,showpacs,groupedaddress]{revtex4}

\usepackage{graphicx}
\usepackage{dcolumn}
\usepackage{bm}
\usepackage{hyperref}
\usepackage{ulem}

\begin{document}

\title{Re-entrant ordering of solute in a colloidal suspension during solvent evaporation}
\author{Sumanta Mukherjee$^{1}$, Arnab Saha$^{2}$, Pralay K. Santra$^{1}$, Surajit Sengupta$^{2,3}$, D. D. Sarma$^{1,*}$}
\affiliation{$^1$ Solid State and Structural Chemistry Unit, Indian Institute of Science, Bangalore 560012, India\\$^{2}$Advanced Materials Research Unit, S. N. Bose National Centre for Basic Sciences, Salt Lake, Kolkata 700091, India\\$^{3}$Centre for Advanced Materials, Indian Association for the Cultivation of Science, Jadavpur, Kolkata 700032 , India}

\date{\today}

\begin{abstract}
We study the phenomenon of self-assembly of silica micro-spheres on a glass plate during evaporation of the solvent from a colloidal suspension. Our experiments unveil an interesting competition between ordering and compaction in a strongly driven, out of equilibrium system arising from a slowing down of dynamics due to an impending glass transition. A suitable choice of experimental conditions minimizing the influence of many other competing phenomena that usually complicate probing of this underlying physics is crucial for our study. A re-entrant behavior in the order-disorder phase diagram as a function of particle density and drying time is established and the results are explained with the help of simulations and phenomenological theory.
\end{abstract}

\pacs{Valid PACS appear here}

\maketitle

Self-assembly of colloidal particles into well-ordered arrays is interesting both due to the diverse technological possibilities\cite{col-assembl,col-assembl-2} and because it involves ordering in systems driven far from equilibrium\cite{non-eq}. Ordered, two-dimensional particle arrays are often produced by evaporating a drop of colloidal suspension on a substrate\cite{evap1,evap2}. The physics of ordering upon evaporation is poorly understood because of the complex interaction of many forces involved viz. lateral capillary forces, convection and floatation, as well as colloid-colloid interactions, and contact line pinning\cite{evap-approx,marangoni,degan,nagayama,nagayama1,jpcb}. In this Letter, we report an experimental realization that enables us to study the underlying interesting physics, including reentrant behavior of such driven systems as a function of just two easily controllable parameters, namely the initial particle density, $\rho_0$ and the drying time, $t_d$, of the solvent, the ordering mechanism being essentially reduced to a competition between drying and slowing down of dynamics due to an impending glass transition\cite{VF}.

\begin{figure*}[t]
\begin{center}
\includegraphics[width=1.0\textwidth]{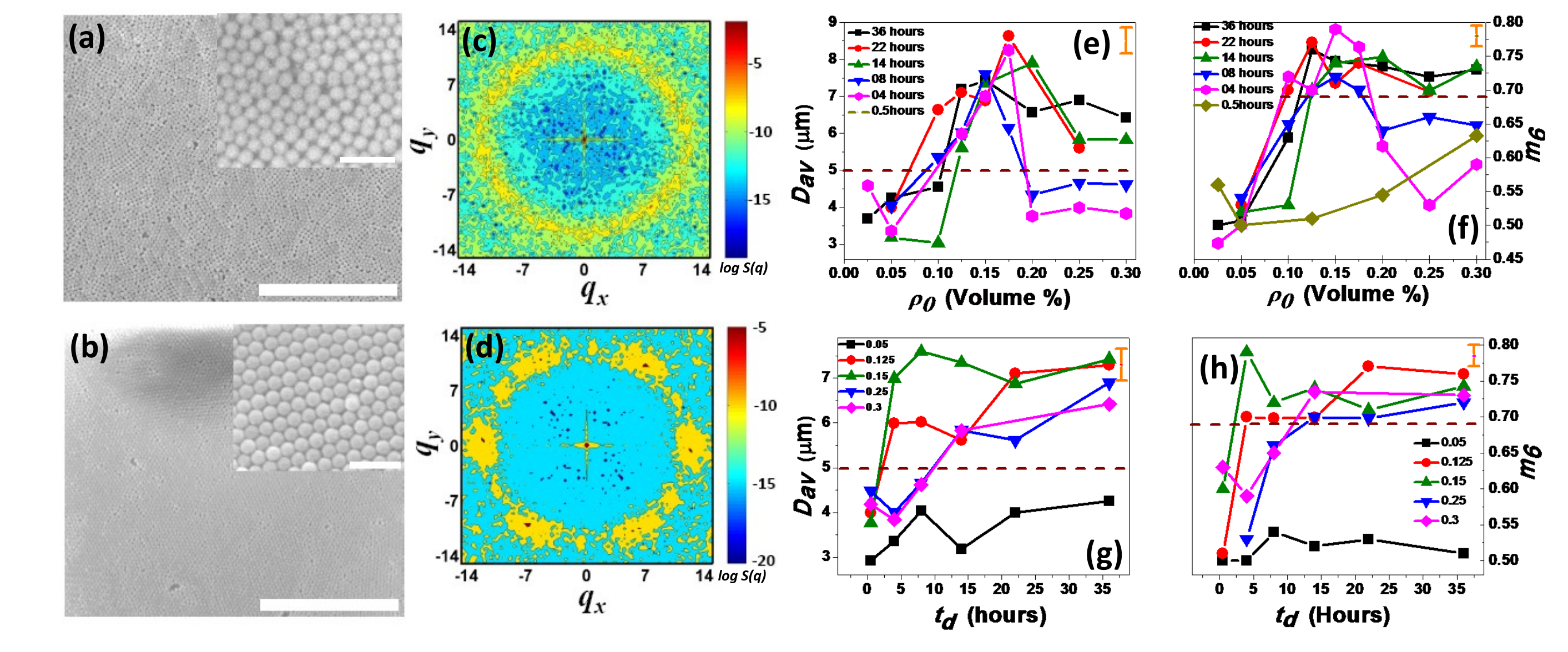}
\caption {(a)and(b): Real space SEM image of silica sphere, Initial volume fraction = 0.15\%, (scale bar = 20 $\mu$m) with a drying time of 30 min and a drying time of 14 hours, respectively. Inset shows the higher magnification image (scale bar 2 $\mu$m) (c) and (d): Corresponding structure factors.
(e) and (f): Change in the average domain size $D_{av}$ and bond order parameter $m_6(\rho_0,t_d)$, respectively, as a function of different silica concentration, $\rho_0$ at different drying time $t_d$. (g) and (h): Change in the average domain size $D_{av}$ and $m_6(\rho_0,t_d)$ as a function of $t_d$ at different $\rho_0$ respectively. a typical error bar in term of the standard deviation calculated from different runs of the measurements is shown as a bar in each of fig (e) to (h)}

\label{fig1}
\end{center}
\end{figure*}

It is commonly believed that drying assisted assembly of colloidal particles proceeds by a two-stage mechanism. In the {\it pinned} state, the contact line at the periphery of the drop is pinned by the substrate, the drop area is fixed and a flow, which advects particles to the periphery, is set up to replenish the liquid lost due to evaporation\cite{degan,nagayama} which occurs primarily at the periphery\cite{evap-approx} due to the increased local curvature. The build-up of particles at the periphery causes self-assembly due to attractive capillary forces, that has been investigated in the past \cite{nagayama,nagayama1}. However, the contact line is eventually depinned and drying recommences when the drop area decreases \cite{jpcb}. The phenomenon of self assembly in this regime has not been investigated so far. The lifetime of the droplet in the two regimes and, consequently, their relative importance depend crucially on the contact angle $\theta_c$ between the drop and the substrate. For small $\theta_c$, the lifetime of the pinned stage is small. We have used this fact to our advantage by decreasing $\theta_c$ below 25$^{\circ}$ by preparing highly hydrophilic substrates, enabling the suspension to dry predominantly under de-pinned conditions. Additionally, because of the low value of $\theta_c$, the evaporation rate from the periphery is weak making the outward flow rate of the liquid and hence the centrifugal particle flux negligible. In order to ascertain the irrelevance of hydrodynamic effects, such as Marongoni flows\cite{marangoni}, in determining the final configuration of the silica particles, we estimate the Peclet number, $P_n = r_p v_p/D_c$, to be less than 0.02 in all cases, where, $r_p$ is the radius of the particle moving with a velocity of $v_p$ and $D_c$ is the diffusion constant for all $t_d$ and $\rho_0$ that we have experimentally investigated.

Colloidal suspension of silica micro spheres of radius $0.554 (\pm 0.023)$ $\mu$m were obtained from Bang's Laboratory, USA. In order to prepare substrate with a highly hydrophilic nature glass slides (cover slips) were heated at 70$^{\circ}$C for 6 hours in a piranha solution ($3:1$ mixture of conc. H$_2$SO$_4$ and H$_2$O$_2$). Then they were cooled, washed several times with de-ionized water and dried. In each set of experiments freshly prepared glass slides were used. Slides obtained by this method are highly hydrophilic with a contact angle less than  25$^{\circ}$ in all cases. In a typical experiment, 60 $\mu$l of silica suspension in deionized water with different silica concentrations (vol$\%$), $\rho_0$, each time was drop-cast on a substrate and dried upto a total drying time, $t_d$, till all trace of the solvent evaporated. Humidity and temperature are the available variables to obtain different values of $t_d$ over a wide range. The temperature was 25 K for all measurements, except when we needed to achieve the shortest $t_d$ (30 mins.) used in our experiments, when the temperature was set at 35 K. It is known that treatment of glass slide with piranha solution makes the surface of the glass substrate negatively charged. Zeta potential measurements show that the surfaces of these colloidal silica particles are also negatively charged (Zeta potential -41.3 $mV$). Therefore, the interaction between the substrate and the particles is repulsive, reducing random pinning of particles on the substrate allowing us to study a cleaner limit of the self-assembly under a driven condition. The interaction between the particles depends not only on their surface charges, but also on the ionic strength of the solvent, which screens the surface charge thus playing an important role in controlling  the interaction between particles and hence the ordering process\cite{tata}. The screening length of the solvent was kept fixed for all the runs, thereby reducing one more controlling parameter in the problem, by fixing the ionic strength of the solvent by the addition of a known concentration of KCl ($10\times^{-5}$ mols/lt) in each case. The random percentage fluctuation of the screening length could thus be reduced far bellow what can be achieved with de-ionized mili-Q water (resistance: 18 Mohm).

Scanning Electron Microscope (SEM) pictures  were taken of the final structure of the colloidal particles after the completion of the drying process using a FEI-Quanta 200. Two representative images are shown in (Fig.1 (a) and (b)). To quantify the extent of ordering, coordinates of every particle in a snap shot of the SEM image over a large area ($50\, \mu m \times 50 \,\mu m$) were obtained using an image digitizer software (Image J).  We characterize each such configuration by calculating the structure factor $S(q) = N^{-1} \sum_{m,n} \exp (i\,{\bf q} \cdot {\bf r}_{mn})$, with the sum taken over all pairs of particles index by $m$ and $n$ and separated by distance ${\bf r}_{mn} = {\bf r}_m - {\bf r}_n$. $S(q)$ is isotropic for disordered configurations and shows crystalline peaks for ordered ones (Fig.1 (c),(d)). The average domain size, $D_{av}$ may be obtained from $S(q)$ using the well known scherrer formula\cite{scherrer}. The orientational (Nelson-Halperin)\cite{Nel-Hal} order parameter  $m_6 = N^{-1} z | \sum_{k=1}^{N} \sum_{n=1}^{z} \exp(i\,6 \theta_{k n})|$, where $N$ is the total number of particles and $z$ the coordination number, was also computed to characterize the final configurations. (Fig.1 (e),(f)) show the variation of $D_{av}$ and $m_6$ with $\rho_0$ for various fixed drying times, $t_d$. It is clear from the results that for slow drying, there is a jump in both $D_{av}$ and $m_6$.  Whereas, in case of faster drying no such transition is observed in either of the two parameters characterizing the extent of ordering, with increase in the concentration of the initial suspension. Similar results are also obtained when the data are plotted against $t_d$ at different $\rho_0$ (Fig.1 (g),(h)).

Guided by the jump, we use threshold value of 5 $\mu$m and 0.69 for $D_{av}$ and $m_6$ respectively, (marked by horizontal lines in each of  (Fig.1 (e)to(h))) to distinguish between ordered and disordered phases, allowing us to obtain a dynamical phase diagram in the $t_d - \rho_0$ space for this whole process (Fig.2). We note that results presented here are not critically dependent on the exact choice of the thresholds and a range of values for the thresholds leads to same results. (Fig.2) clearly shows that the system is invariably disordered for $\rho_0 < 0.1\%$ of silica particle. For $\rho_0 > 0.1\%$, we find that the system is disordered for small $t_d$ and is ordered for larger $t_d$; the critical $t_d$ to bring about the ordered structure increases somewhat for a higher $\rho_0$. Thus the data clearly establish a disorder to order transition as a function of $\rho_0$ across 0.1$\%$ for a wide range of drying time $t_d \geq$ 4 hrs. Likewise, there is a disordered to ordered transition as a function of the increasing drying time $t_d$ for $\rho_0 >$ 0.1$\%$. Interestingly, present data also show an ordered to disordered transition on increasing $\rho_0$ for lower values of $t_d \leq$ 10 hrs. The data points shown in (Fig.2 ) therefore establish a dynamical phase diagram with a re-entrant disorder $\to$ order $\to$ disorder transition in this system, the phase boundary going through a minimum showing that it is possible to optimize both $t_d$ and $\rho_0$ in order to obtain large, ordered, self-assembled arrays.

\begin{figure}[h]
\begin{center}
\includegraphics[width=0.5\textwidth]{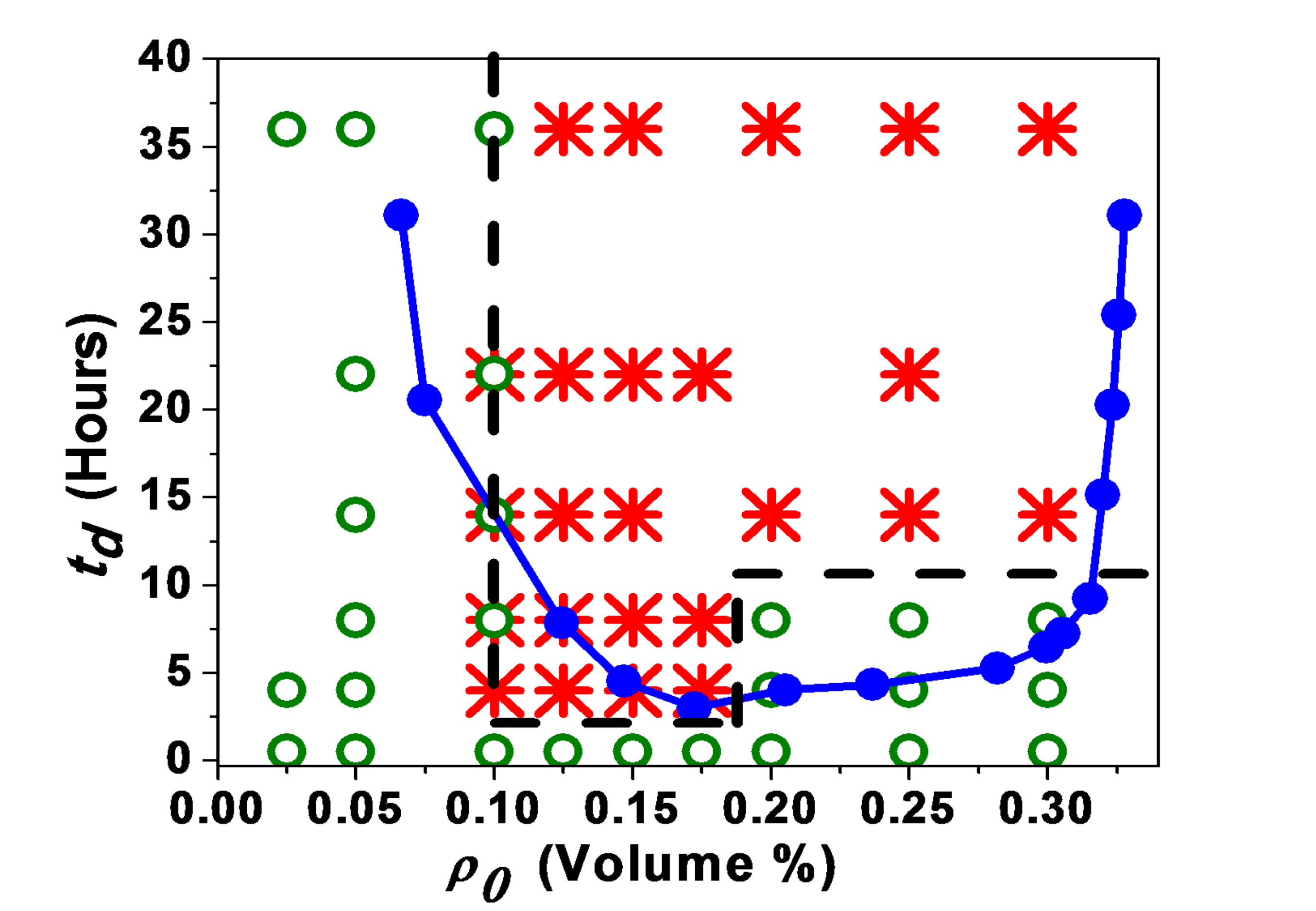}
\caption{The experimentally obtained dynamical phase diagram in the $t_d - \rho_0$ plane. The red asterisk and the green open circle represent ordered and disordered configurations, respectively. Overlapping asterisks and circles denote points at the phase boundary, where independent experimental runs yielding contrasting extent of ordering across the threshold, indicative of the existence of the phase boundary. Also shown are results from our BD simulations: blue circles. We use $4000$ particles for our simulations and a maximum of $625$ cells each of size $5$ to represent the solvent field $\phi$.  A time step of $10^{-3}$ is used to update both particle positions and $\phi$. Comparing our numerical solution with the analytic result for Ising droplets, we estimate our accuracy for $\phi$ updates to be $1$ in $10^{-6}$. In order to compare our results with experiments, we have multiplied the axes $\rho_0\kappa^2$ by $0.2$ and $\gamma t_d $ by $1.5$ to obtain $\rho_0$ in units of Vol. \% of silica and $t_d$ in hours. The scale factor for $\rho_0$ was obtained from an independent measurement of $\kappa$ for the silica particles in solution while $t_d$ was simply scaled to obtain best fit with experiments.}
\label{simres}
\end{center}
\end{figure}

To understand the effect of  $t_d$ and  $\rho_0$ on the final structure, we model\cite{simul} the drying process as a 2d liquid-gas phase transition. The local density of the solvent $\phi({{\bf} r},t)$ at position ${{\bf}r}$, satisfies\cite{chaikin}, $\dot \phi = \delta {\cal F} /\delta \phi$ with ${\cal F}[ \lbrace \phi \rbrace ] = \int d^2{\bf r}  \,\,a \phi^2 + \phi^4 + c (\nabla \phi)^2 + \mu_s \phi,$ where $\mu_s$ is the chemical potential of the solvent controlling the rate of drying and $a,c$ are parameters. The homogeneous part of the free energy density has minima at $\phi_{0}=\pm\sqrt{-2 a}/2$ when $a<0$ and $\mu_s=0$ denoting the liquid $\phi=\phi_{0}$ and vapour $\phi=-\phi_{0}$ phases of the solvent. The particle dynamics is modeled using an over-damped Brownian dynamics (BD) scheme, $ \Gamma (\phi) \dot {\bf r}_i =  F_i $, where the force $F_i = \sum_j \frac{\partial V_{ij}}{\partial {\bf r}_{ij}} +\frac{\partial \phi}{\partial {\bf r}_i}+\eta_i$ consists of three parts (1) from the pair interaction $V_{ij} = V_0\exp(-\kappa r_{ij})/r_{ij}$ (screened Coulomb) between the particles $i$ and $j$, (2) from surface tension forces produced at the contact line and (3) Gaussian white noise $\eta_i$ at the ambient temperature. In order to model the fact that the silica particles are immobile without the solvent, we take the mobility $\Gamma = (\tanh(\gamma \phi)+1)/2$, which vanishes in the limit $\phi \to -\phi_0$ \footnote[1]{Note that the fluctuation dissipation theorem is satisfied only within the solvent ($\phi\rightarrow \phi_0$).}. The prefactor, $V_0$, sets the energy and $\kappa$ the screening length the scale for distances. The control parameters are the initial density of the particles in the drop, $\rho_0$, and the drying rate which is controlled by varying $\mu_s$.

For a given  $\rho_0$, we prepare an initial circular drop ($\phi ({\bf r},t=0) = \phi_0$ for $|{\bf r}| = R$  and $-\phi_0$ otherwise) containing randomly dispersed particles. As the drop evaporates, the shrinking circular contact line ($\phi({\bf r},t) = 0$) draws the particles together increasing the particle concentration. When $\phi = -\phi_0$ everywhere all dynamics ceases ($\Gamma(-\phi_0) = 0$). In order to make a direct comparison with experimental results we calculate $m_6$, taking $m_6 = 0.69$ as the threshold for ordering. The resulting phase line is shown in (Fig.2). Considering the simple model that we have used, the agreement between the calculated phase line and the one suggested by the experimental results, including re-entrance, is remarkable.

\begin{figure}[t]
\begin{center}
\includegraphics[width=0.5\textwidth]{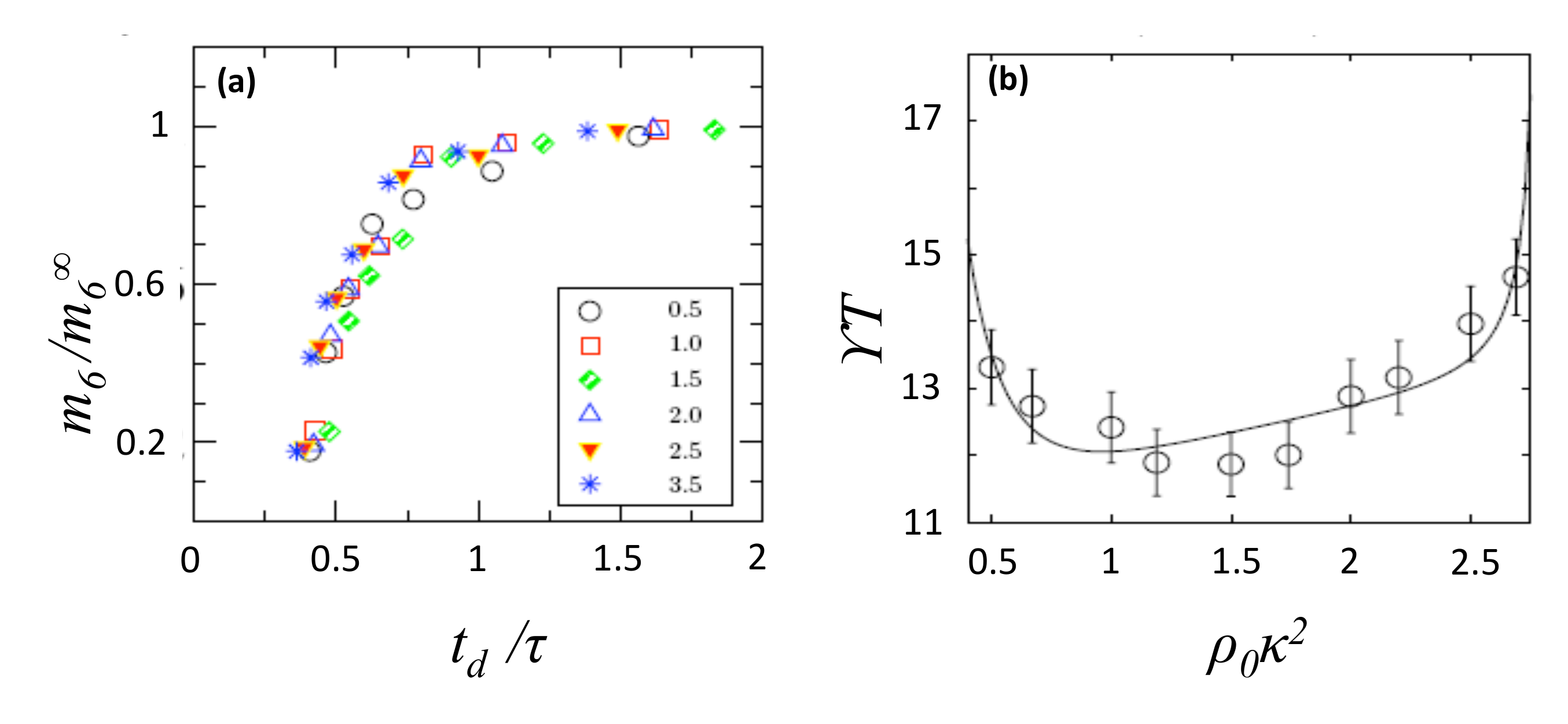}
\caption{(a) Data collapse plot of the bond order parameter $m_6/m_6^{\infty}$ vs $t_d/\tau$ for various values of $t_d$ and $\rho_0$; $m_6^{\infty}$ is the saturation value of $m_6$ and $\tau$ is the ($\rho_0$ dependent) relaxation time. (b) Plot of the relaxation time $\tau$ vs $\rho_0 \kappa^2$ showing a minimum. Open circles represent data from simulations as obtained from the scaling plot in (b), where the line is a fit with the analytic form (see text).(c) Plot of the average coordination number $z$ (open squares, scale to the left) and fractional increase in the density $\delta \rho/\rho_0$ (filled squares, scale to the right) as a function of $\rho_0 \kappa^2$ for fixed $t_d = 7$ hours.}
\label{insight}
\end{center}
\end{figure}

The average order parameter scales as $m_{6}(\rho = m_6^{\infty} f (t_d/\tau)$ , where $m_6^{\infty}$, the saturation value and the relaxation time $\tau$ are functions of $\rho_0$ alone, as illustrate by the collapse of the simulated $m_{6}/ m_6^{\infty}$ as a function of $(t_d/\tau)$ for several $\rho_0$ values shown in (Fig. 3(a)). It is clear  that $\tau$ itself goes through a minimum (see Fig. 3(b)). The reentrant behavior as seen in the dynamical phase diagram is thus coded into the shape of $\tau(\rho_0)$. Indeed, the dynamical phase diagram is a plot of  the contour $m_{6} =\, $constant ($=0.69$ in our case) being given by $t_d = \tau(\rho_0) \times f^{-1}(m_6)$.

There are two timescales associated with the relaxation of the order parameter $\tau = \tau_1 + \tau_2$. The time, $\tau_1$,  required for the concentration to increase from the initial value $\rho_0$ to the level $\rho_f$ the freezing density,  can be easily estimated\cite{grant}, being simply given by the time taken by an Ising droplet to shrink by a fixed amount viz. $\tau_1 = A_1 \rho_0^{-1} + A_2 \rho_0^{-3/2} + A_3$ where $A_i$ depend on the parameters $c$ and $\mu_s$ as well as $\rho_f$. Note that $\tau_1$ diverges as  $\rho_0 \to 0$. The time required for order to develop, $\tau_2$, is the time needed by the system for exploring the space of all configurations.  A Vogel-Fulcher like empirical form\cite{VF} for the relaxation time, viz. $\tau_2 = \tau_0 \exp[ C/(\rho_J - \rho_0)]$ with $\tau_0$, $C$ and $\rho_J$ as free parameters, fits our simulation data rather well (Fig.3(b)). For small $\rho_0$, fluctuations allow the system ample time to explore configuration space during the drying process, though one requires a large drying time ($t_d \langle \tau_1$) in order to raise the density of the solution to the value needed to initiate ordering. On the other hand if $\rho_0$ is high, rapid compaction causes the particles to jam, setting up a network of force chains\cite{cates,liu,slow-relax} which resists any further increase of density. Ordering in this limit is dominated by $\tau_2$. Slowing down of the dynamics is caused by divergences of $\tau_1$ and $\tau_2$, alternately at the low and high density limit, causing the reentrant transition observed in these experiments. Deep inside the high density region an additional complication, viz. the formation of multiple layers commences in the experimental system cutting off the divergence of $\tau_2$.

\noindent
In this Letter, we have shown that drying induced ordering in carefully prepared samples can be used to study in detail the competition between the thermal and athermal processes of ordering and compaction. Minimizing additional complications that often arise in most experimental systems, eg. multilayer formation, hydrodynamic effects, pinning and poly-dispersity, which can be investigated separately in the future,  our experiments, nonetheless, show a rich phase diagram, including a clear sign of reentrant behavior as a consequence of this competition at the simplest level. We explain these phenomena in this system driven far from equilibrium as arising from a slowing down of dynamics due to an impending glass transition in terms of simple physical arguments based on detailed Brownian dynamics simulations.

We thank Bulbul Chakraborty for useful discussions. The authors thank the Department of Science and Technology, Government of India, for financial support. DDS acknowledges the J C Bose Fellowship.
\vskip 1 cm
\noindent
* Also at JNCASR, Bangalore 560054, (sarma{@}sscu.iisc.ernet.in)


\begin{thebibliography}{}

\bibitem{col-assembl} Mirkin, C. A. {\&} Rogers J. A.(ed.) Special issue on Emerging Methods for Micro and Nano fabrication , {\it MRS Bull.} {\bf 26}, 506 (2001).

\bibitem{col-assembl-2} Y. Xia, B. Gates, Y. Yin and Y. Lu, Adv. Mater. {\bf 12}, 693 (2000).

\bibitem{non-eq} E Rabani, D. R. reichman, P. L. Geissler and L. E. Brus, Nature {\bf 426}, 271 (2003).

\bibitem{evap1} T. Brezesinski, M. Groenewolt, A. Gibaud, N. Pinna, M. Antonietti and B. M. Smarsly, Adv. Mater. {\bf 18}, 2260 (2006).

\bibitem{evap2} S. Rakers, L. F. Chi, and H. Fuchs, Langmuir {\bf 13}, 7121 (1997)

\bibitem{marangoni} R. Bhardwaj, X. Fang and D. Attinger, New J. Phys. {\bf 11}, 075020 (2009).

\bibitem{degan} R. D. Deegan, O. Bakajin, T. F. Dupont, G. Huber, S. R. Nagel,  and T. A. Witten, Nature {\bf 389}, 827 (1997).

\bibitem{nagayama} Q. Yan, L. Gao, V. Sharma, Y. Chiang and C. C. Wong, Langmuir {\bf 24}, 11518 (2008).

\bibitem{evap-approx} H. Hu and R. G. Larson, J. Phys. Chem. B {\bf 106}, 1334 (2002).

\bibitem{nagayama1} N. D. Denkov, O. D. Velev, P. A. Kralchevsky, I. B. Ivanov, H. Yoshimura and K. Nagayama, Langmuir {\bf 8}, 3183 (1992).

\bibitem{jpcb} D. M. Soolaman, and H. Yu, J. Phys. Chem. B {\bf 109}, 17967 (2005).

\bibitem{VF} C. A. Angell, Science {\bf 267}, 1924 (1995).

\bibitem{tata} P. S. Mohanti and B. V. R. Tata, J. Colloid Interface Sci. {\bf 264}, 101 (2003).

\bibitem{scherrer} B. D. Cullity and S. R. Stock, {\it Elements of X-ray-diffraction}, (Addison-Wesley, New Jersey, U.S.A. 2001).

\bibitem{Nel-Hal} D. R. Nelson and B. I. Halperin, Phys. Rev. B {\bf 19}, 2457 (1979).

\bibitem{simul} D. Frenkel and B. Smit, {\it Understanding Molecular Simulations},(Academic Press, San Diego, U.S.A. 2002).

\bibitem{chaikin} P. M. Chaikin and T. C. Lubensky, {\it Principles of Condensed Matter Physics}, (Cambridge University Press, NY, 1995).

\bibitem{grant} Godreche C. (ed.) {\it Solids far from Equilibrium} (Cambridge University Press, Cambridge, 1992)

\bibitem{cates} M. E. Cates, J. P. Wittmer, J.P. Bouchaud and P. Claudin, Phys. Rev. Lett. {\bf 81}, 1841 (1998).

\bibitem{slow-relax} P. Richard, M. Nicodemi, R. Delannay, P. Ribi\`ere and D. Bideau, Nature Mater. {\bf 4}, 121 (2005).

\bibitem{liu}  A. J. Liu and S. R. Nagel, Nature {\bf 396}, 21 (1998).





\end{thebibliography}
\end{document}